\documentclass[CEJP,DVI]{cej}
\usepackage{layout}
\usepackage[OT4]{fontenc}
\usepackage{amsmath}
\usepackage{textcomp}
\usepackage{hyperref}
\usepackage{amsmath}
\usepackage{amssymb}

\title{Transverse ultrasonic anomaly in $La_{1/3}$$Sr_{2/3}$$MnO_{3}$}

\articletype{Research Article}

\author{Kong Hui\email{konghui@ahut.edu.cn},
        Tong lianhai, Zhou yahui}

\institute{
     \inst{1} School of Metallurgy Engineering, Anhui University of Technology\\ Maanshan, Anhui, 243002, P R China\\
         \inst{2} School of Metallurgy Engineering, Anhui University of Technology\\ Maanshan, Anhui, 243002, P R China\\ \inst{3} School of Metallurgy Engineering, Anhui University of Technology\\ Maanshan, Anhui, 243002, P R China\\  }

\abstract{The charge ordering (CO) transition in polycrystalline $La_{1/3}$$Sr_{2/3}$$MnO_{3}$ has been studied by measuring the resistivity, magnetization and transverse ultrasonic velocity. At about 235K, a conspicuous increase in resistivity was observed, while the magnetization shows a cusp structure, corresponding to an antiferromagnetic charge ordering transition. Around this transition temperature, dramatic anomaly in transverse sound velocity was observed. The simultaneous occurrence of electron, magnon and phonon anomalous features implies strong spin-phonon coupling and electron-phonon in $La_{1/3}$$Sr_{2/3}$$MnO_{3}$. The analysis suggests that the spin-phonon interaction is due to single-ion magnetostriction, and electron-phonon coupling originates from the Jahn-Teller effect of $Mn^{3+}$.}


\begin{document}
\maketitle


\section{Introduction}

\setlength{\parindent}{2em}Hole-doped perovskite manganites $R_{1-x}$$A_{x}$$MnO_{3}$ (R: trivalent ions, A: divalent alkaline-earth ions) have been a subject of intense investigations because of their special structural, magnetic, and electronic properties, such as colossal magnetoresistance effect, phase separation and charge ordering transition [1-4]. Up to now, it has been widely recognized that these unique physical properties is a consequence of the coupling or the competition among the degrees of freedom of charge, spin, lattice and orbit.\par
Among different manganites, $La_{1-x}$$Sr_{x}$$MnO_{3}$ system has been in-depth studied since it presents a very complex phase diagram, especially at low Sr-density. The unique coexistence of ferromagnetism and insulting behavior in low doping range (x=0.11-0.14) highlights its significance beyond the double-exchange (DE) mechanism [5-8]. For example, $La_{1-x}$$Sr_{x}$$MnO_{3}$ (x=0.13) sample first shows a cooperative Jahn-Teller distortion at $T_{JT}$$\approx$265K, then undergoes a ferromagnetic transition at $T_{C}$$\approx$190K, and finally transforms into ferromagnetic insulating state at $T_{CO}$$\approx$163K. $^{55}$Mn NMR measurements suggest that its ground state is composed of three different spin and orbital entities [8]. For $La_{0.7}$$Sr_{0.3}$$MnO_{3}$, it shows a maximum alternating current magnetoresistance of -34\% at $T_{C}$ in a field of 600Oe [9]. In 2012, Barilk et.al reported four probe ac electrical impedance in $La_{0.7}$$Sr_{0.3}$$Mn_{1-x}$$Fe_{x}$$O_{3}$ (x$\leq$0.15) as a function of temperature and magnetic field in response to radio frequency (f=0.1-5MHz) ac current flowing directly through the sample [10]. For $La_{0.5}$$Sr_{0.5}$$MnO_{3}$, Kallel et.al pointed out that it exhibits interesting behaviors of charge-ordering, ferromagnetic states and a good conductivity down to the lowest temperatures [11]. However, the high-doping regime of $La_{1-x}$$Sr_{x}$$MnO_{3}$ (x$\sim$2/3) has been less intensively studied though in this range, charge ordering (CO) accompanies antiferromagnetic (AFM) spin ordering [12].\par
As a sensitive tool, ultrasonic technique has proven to be successful for studying electron-phonon and spin-phonon couplings in perovskite materials [13-14]. It has been found that around the $T_{CO}$, the abnormal change of velocity is closely related to the strong electron-phonon coupling. In 2000, ultrasonic measurement has been applied in the perovskite manganite $La_{1-x}$$Sr_{x}$$MnO_{3}$ (x=0.12, 0.165, 0.3) with both orbital and charge degrees of freedom [15]. Results shows that the electron-lattice interaction gives rise to a structural change accompanied by the orderings of the orbit and charge degrees of freedom. Moreover, the temperature dependence of longitudinal ultrasonic velocity has been measured in $La_{1/3}$$Sr_{2/3}$$MnO_{3}$ [14]. To complete the ultrasonic characterization of $La_{1/3}$$Sr_{2/3}$$MnO_{3}$, we perform transverse ultrasonic measurement in order to understand the microscopic origin of the CO transition in $La_{1/3}$$Sr_{2/3}$$MnO_{3}$.

\section{Experimental procedure}

The polycrystalline $La_{1/3}$$Sr_{2/3}$$MnO_{3}$ was synthesized by the chemical citrate-gel route. Stoichiometric amounts of aqueous metal nitrates were prepared by dissolving $La_{2}$$O_{3}$, SrC$O_{3}$ and Mn(N$O_{3}$)$_{2}$ in a minimum amount of concentrated nitric acid. After adding the citric acid, the pH value of solution is adjusted to 6$\sim$7. The solution was subsequently stirred and evaporated at 80$^{\circ}$C to form the polymeric precursor, which was ignited in air to remove the organic contents. The resulting ash was ground to fine powders and calcined at 800$^{\circ}$C in air for 10 h. The final obtained powder was pressed into pellets at 300MPa and then sintered at 1500$^{\circ}$Cin air for 20h, and cooled to room temperature at a rate of 1$^{\circ}$C $min^{-1}$.\par
The crystal structure was characterized using a Japan Rigaku MAX-RD powder X-ray diffractometer with Cu K $\alpha$ radiation ($\lambda$=1.5418$\r{A}$ ). The resistivity was measured by the standard four-probe technique. The zero-field-cooled magnetization was measured in an external magnetic field of 100Oe by a commercial quantum device (SQUID; Quantum Design MPMSXL).\par
The specimen for ultrasonic measurement was in the form of flat disk, 4.54mm thick, and was hand-lapped to a parallelism of faces better than 2 parts in $10^{4}$. The transverse ultrasonic velocity measurement was performed on the Matec-7700 oscillator/receiver series with a conventional pulsed echo technique. Transverse ultrasonic wave pulses are generated by a Y cut quartz transducer at a frequency of 10 MHz. It was bonded to the sample surface with nonaqueous stopcock grease and converted the energy to ultrasound. The absolute percentage error in velocity measurement is $\pm$0.1\%, and the relative error is $10^{-6}$. All experiments were taken in a closed-cycle refrigerator during the warm-up from 20 to 300K at the rate of about 0.25K/min.

\section{Experimental results and discussion}

The X-ray diffraction pattern of $La_{1/3}$$Sr_{2/3}$$MnO_{3}$ is shown in Fig.1. The sample is of single phase without detectable secondary phase. The diffraction peaks can be indexed based on the pseudo-cubic structure.\par
The temperature dependences of resistivity and magnetization for $La_{1/3}$$Sr_{2/3}$$MnO_{3}$ are displayed in Fig.2. The resistivity shows semiconductive behavior over the entire temperature range. And a conspicuous increase in resistivity was observed near the charge ordering temperature $T_{CO}$, while the magnetization shows a cusp structure. This feature implies the onset of the charge ordering state, which is composed by the $Mn^{3+}$ and $Mn^{4+}$.\par
Fig.3 shows the temperature dependence of the transverse ultrasonic velocity ($V_{T}$) for $La_{1/3}$$Sr_{2/3}$$MnO_{3}$ at a frequency of 10 MHz. Since the transverse sound only propagate along the solid state and nonaqueous stopcock grease freezes at about 250K, the V can only be measured up to 250K. The $V_{T}$ softens conspicuously as the temperature decreasing, and there is a minimum around $T_{CO}$. Below $T_{CO}$, the $V_{T}$ stiffens dramatically and iys relative stiffening is more than 30\%.\par
It can be seen that the ultrasonic anomalies occur around $T_{CO}$, and accompany with the conspicuous decreasing of magnetization and sharp increasing of resistivity. This clearly shows the existences of spin-phonon interaction and electron-phonon interaction. The detailed discussions are shown below.\par
For the spin-phonon interaction, it is due to magnetostriction. The thermal fluctuations of the spins become large near the magnetic transition point and may produce an appreciable scattering of the phonons, resulting in a significant coupling between the spins and the phonons. This coupling is classified into two types [16]. If the coupling is due to volume magnetostriction, longitudinal velocity should be anomalous and transverse velocity should not. On the other hand, for single-ion (linear) magnetostriction, both transverse and longitudinal modes should show anomalies. The similar ultrasonic anomalies in transverse and longitudinal modes in our sample (shown in Fig.3) indicate that the spin-phonon interaction in $La_{1/3}$$Sr_{2/3}$$MnO_{3}$ is due to single-ion magnetostriction rather than volume magnetostriction.\par
However, it should be mentioned that according to the Landau-Khalatnilov theory, AFM ordering usually decreases the elastic stiffness due to the magnetostriction effect, and the relative change in sound velocity is usually less than 0.1\% [17]. For example [18], according the temperature dependence of longitudinal velocity for the similar perovskite manganese oxide (CaMn$O_{3}$), it can be seen that its velocity monotonously increases with the temperature decreasing. And the relative change of velocity around AFM transition temperature is only -0.07\%, which is due to the antiferromagnetic spin fluctuations alone. While in our sample, the relative change is more than 40\%. Thus, the spin-phonon cannot explain the ultrasonic anomalies alone.\par
For the electron-phonon interaction, its existence in perovskite manganese has been widely accepted, and normally considered to originate from the Jahn-Teller effect of $Mn^{3+}$. For $Mn^{3+}$, its electronic structure is \\3$d^{4}$ ($t_{2g}$$^{3}$$e_{g}$$^{1}$ configuration), which means one unpaired electron resides in the two-fold degenerate eg level. When the eg electrons are localized, such as charge ordering transition, this degeneracy may be removed by a Jahn-Teller distortion. This kind of localized lattice distortion is induced to reduce the electrostatic interaction of the electron and the surrounding oxygen atoms, and leads to the ultrasonic anomalies.\par
The upper deduction is based on the solid foundation. On the one side, according to the Jahn-Teller theory, the coupling of the electronic states of the ions to the long-wavelength acoustic phonons will cause one or more elastic constants to undergo an anomalous decrease as the phase transition is approached [19]. This decrease reflects the instability of the lattice. On the other side, the calculation, which based on the Hamiltonian of small polarons with strong electron-phonon coupling, confirmed that CO transition in manganites results in the decreasing of the sound velocity above $T_{CO}$ and the increasing below $T_{CO}$ [20]. These theoretical results are qualitatively similar to our observations. Moreover, electron-phonon coupling can explain the large relative change of ultrasonic velocity. For example, in $La_{0.27}$$Ca_{0.73}$$MnO_{3}$, the relative stiffening change of transverse velocity is about 50\% [21], which is much larger than that of longitudinal mode. Both this value and contrastive phenomenon are similar to that of our sample. Therefore, it is reasonable to attribute the ultrasonic anomalies in our sample to the electron-phonon coupling via the Jahn-Teller effect of $Mn^{3+}$.

\section{Conclusion}

In summary, we have performed the measurements of resistivity, magnetization and transverse ultrasonic velocity in polycrystalline $La_{1/3}$$Sr_{2/3}$$MnO_{3}$. At about 235K ($T_{CO}$), the charge ordering transition occurs. Upon cooling down from high temperature, an obvious softening in velocity above $T_{CO}$ and dramatic stiffening below $T_{CO}$ are observed. The simultaneous occurrence of electron, magnon and phonon anomalous features implies strong spin-phonon coupling and electron-phonon in $La_{1/3}$$Sr_{2/3}$$MnO_{3}$. Through combining with longitudinal ultrasonic results, the analysis indicates that the spin-phonon interaction is due to single-ion magnetostriction rather than volume magnetostriction, and electron-phonon coupling originates from the Jahn-Teller effect of $Mn^{3+}$.

\section*{Acknowledgment}

This work was supported by the National Natural Science Foundation of China (No.11004002).

\section*{Figure Captions:}
\begin{figure}
\includegraphics[height=7cm]{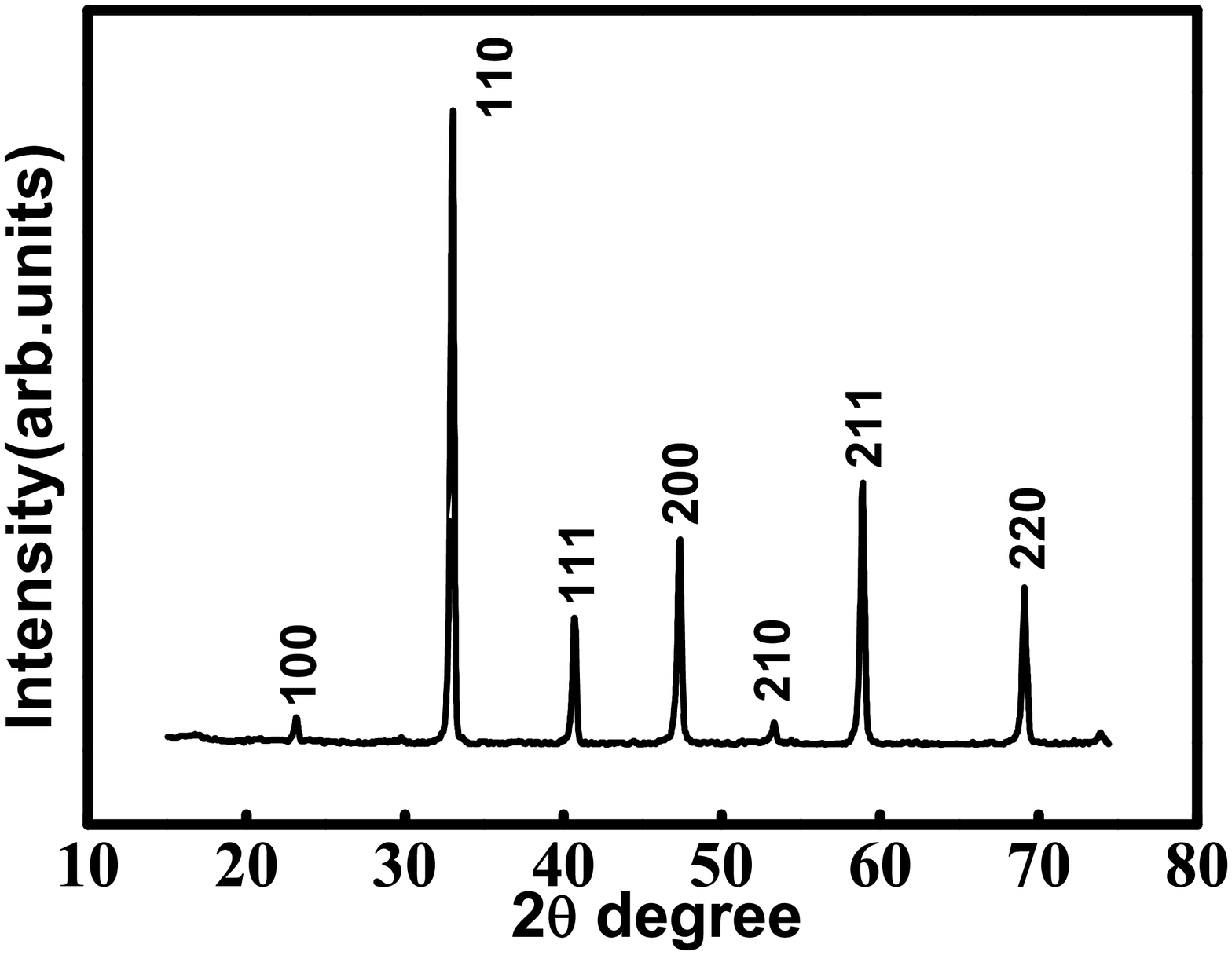}
\caption{XRD pattern of $La_{1/3}$$Sr_{2/3}$$MnO_{3}$ at room temperature}
\includegraphics[height=7cm]{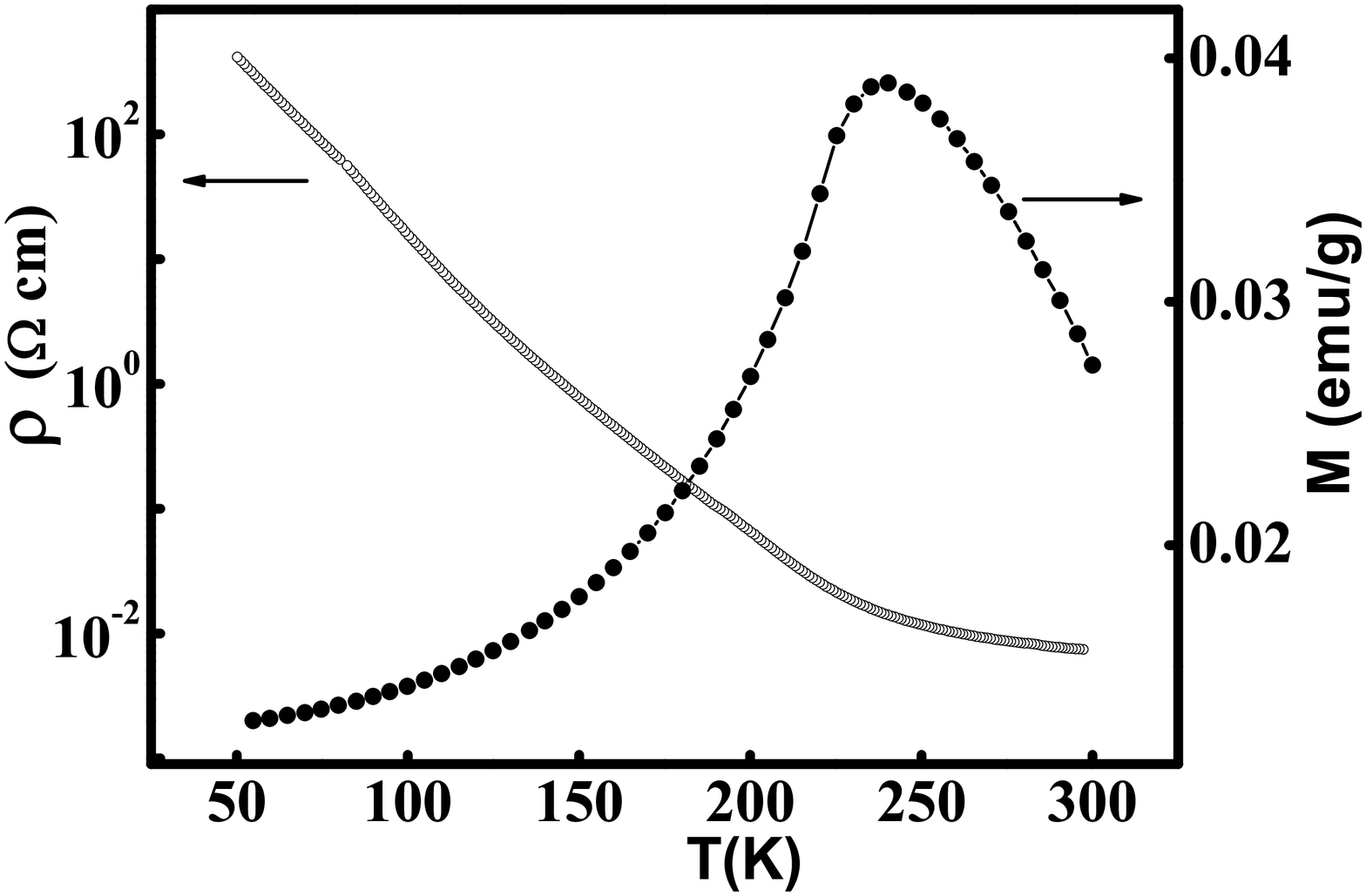}
\caption{Temperature dependences of resistivity and magnetization for $La_{1/3}$$Sr_{2/3}$$MnO_{3}$}
\includegraphics[height=7cm]{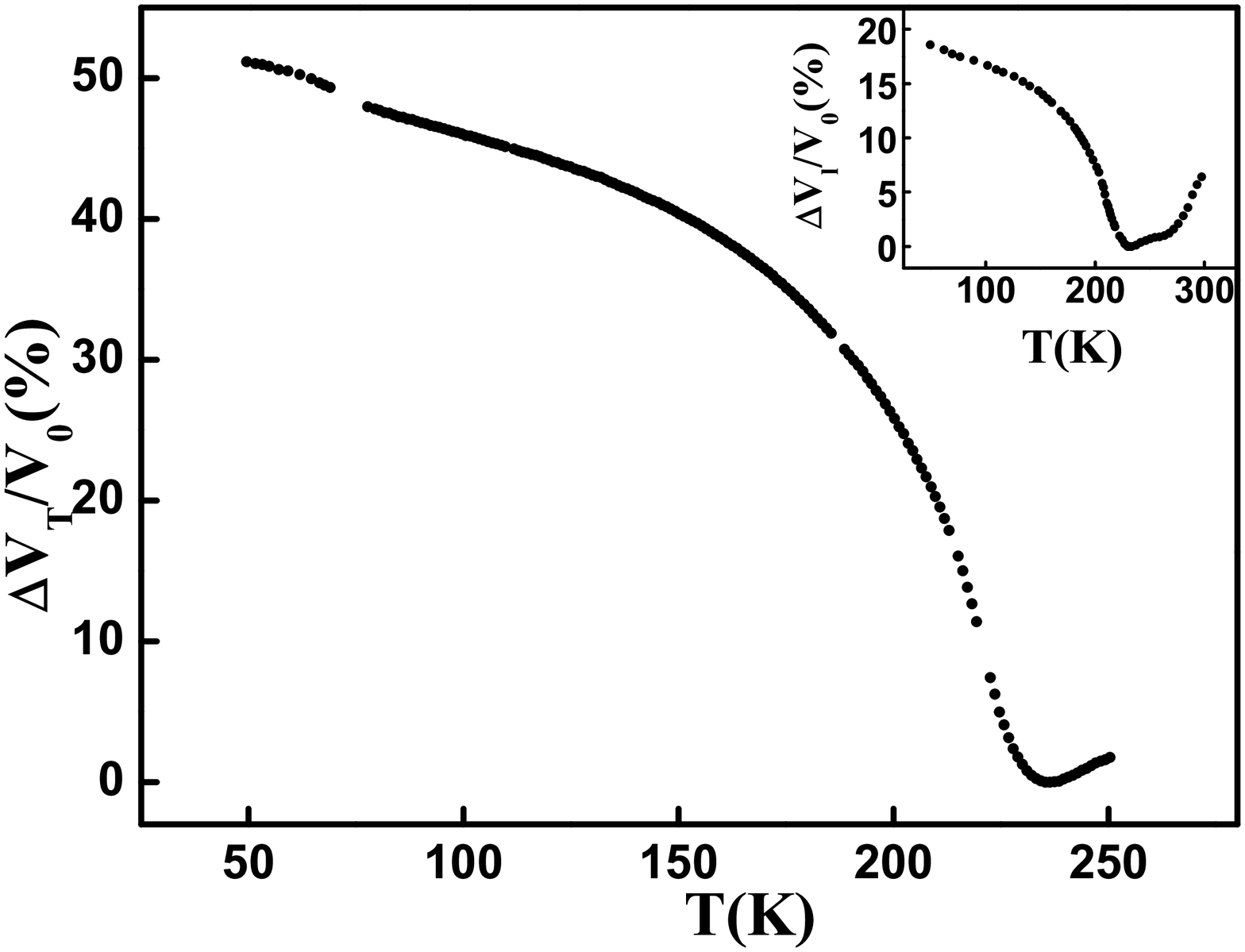}
\caption{Temperature dependence of the transverse ultrasonic velocity ($V_{T}$) for $La_{1/3}$$Sr_{2/3}$$MnO_{3}$}
\end{figure}

fig. 1 XRD pattern of $La_{1/3}$$Sr_{2/3}$$MnO_{3}$ at room temperature

fig. 2 Temperature dependences of resistivity and magnetization for $La_{1/3}$$Sr_{2/3}$$MnO_{3}$

fig. 3 Temperature dependence of the transverse ultrasonic velocity ($V_{T}$) for $La_{1/3}$$Sr_{2/3}$$MnO_{3}$
The inset is the temperature dependence of longitudinal ultrasonic velocity ($V_{1}$) for $La_{1/3}$$Sr_{2/3}$$MnO_{3}$

\end{document}